\documentclass[numreferences]{kluwer}
\newdisplay{guess}{Conjecture}

\def\be{\begin{equation}}
\def\ee{\end{equation}}
\def\bea{\begin{eqnarray}}
\def\eea{\end{eqnarray}}

\def\l{\left}
\def\r{\right}

\begin{document}
\begin{opening}
\title{Exact solutions of the Li\'{e}nard and generalized Li\'{e}nard type ordinary
non-linear differential equations obtained by deforming the phase space coordinates
of the linear harmonic oscillator}
\author{Tiberiu \surname{Harko}\thanks {email: t.harko@ucl.ac.uk}}
\institute{Department of Mathematics, University College London, Gower Street, London
WC1E 6BT, United Kingdom}

\author{Shi--Dong  \surname{Liang}\thanks {email: stslsd@mail.sysu.edu.cn}}
\institute{State Key Laboratory of Optoelectronic Material and Technology, and Guangdong Province Key Laboratory of Display Material and Technology, School of Physics and Engineering,
Sun Yat-Sen University, Guangzhou 510275, People’s Republic of China}
\runningauthor{T. Harko and S. D. Liang}
\runningtitle{Exact solutions of the Li\'{e}nard type differential equation}
\date{\today}

\begin{abstract}
We investigate the connection between the linear harmonic oscillator equation and some classes of  second order nonlinear ordinary differential equations of Li\'{e}nard and generalized Li\'{e}nard type, which physically describe important oscillator systems. By using a method inspired by quantum mechanics, and which consist on the deformation of the phase space coordinates of the harmonic oscillator,  we generalize the equation of motion of the classical linear harmonic oscillator to several classes of strongly non-linear differential equations. The first integrals, and a number of exact solutions of the corresponding equations are explicitly obtained.  The devised method can be further generalized to derive explicit general solutions of nonlinear second order differential equations unrelated to the harmonic oscillator. Applications of the obtained results for the study of the travelling wave solutions of the reaction-convection-diffusion equations, and of the large amplitude free vibrations of a uniform cantilever beam are also presented.

\hspace{-0.5cm}{\bf Math Subject Classification:} 34A05: 34A34: 34A25: 70K40
\end{abstract}
\keywords{Li\'{e}nard equation: harmonic oscillator: phase space deformation: integrability condition: exact solutions}
\end{opening}

\section{Introduction}

An important equation in mathematical physics is the Li\'{e}nard type ordinary  second order nonlinear differential equation of the form \cite{Lien}
\be
\ddot{x}(t)+f(x)\dot{x}(t)+g(x)=0,
\ee
where a dot represents the derivative with respect to the time $t$, and $f$ and $g$ are arbitrary $C^1$ functions of $x$. Together with its generalization, the Levinson-Smith type equation \cite{Lev}
\be
\ddot{x}(t)+f\left(x,\dot{x}\right)\dot{x}(t)+g(x)=0,
\ee
 where $f$ is an arbitrary function of $x$ and $\dot{x}$, the Li\'{e}nard equation plays an important role in many areas of physics, biology and engineering \cite{2}. The mathematical properties of these types of equations have been intensively investigated from both mathematical and physical point of view, and their study remains an active field of research in mathematical physics \cite{lit0} - \cite{lit6}.

One of the major scientific and engineering applications  of the generalized Li\'{e}nard type equations is the mathematical study of a large number of non-linear oscillations, which  can be described by
\be
\ddot{x}+f\left(\dot{x}\right)g(x)+h(x)=0,
\ee
where $f$, $g$ and $h$ are arbitrary $C^1$ functions. In many practical situations the function $f\left(\dot{x}\right)$ can be approximated as
\be
 f\left(\dot{x}\right)=\alpha \dot{x}^3+\beta \dot{x}^2+\gamma \dot{x}+\delta,
 \ee
 where $\alpha $, $\beta $, $\gamma $, $\delta $ are constants \cite{Mic,Nay}. From a physical point of view the Li\'{e}nard equation can be interpreted as the mathematical generalization of the equation of the damped oscillations,
 \be
 \ddot{x}+\gamma \dot{x}+\omega ^2 x=0,
 \ee
 with $\gamma ={\rm constant}$ and $\omega ^2={\rm constant}$, respectively \cite{Ben}. For $\gamma =0$ we obtain the equation of the linear harmonic oscillator, which represents one of the fundamental equations of both classical and quantum physics. Generally, a linear oscillation can be described by the second order ordinary differential equation \cite{Ben}
 \be
 \ddot{x}+f(t)\dot{x}+g(t)x=0.
  \ee
  The second order linear equation describing the oscillations of a complex physical system can be reduced to a Riccati equation \cite{Ha,Ha1,Ha2}. Hence the solutions of the Riccati equation corresponding to the given oscillator can be used to study the linear oscillations of classical systems. By using N. Euler's theorem on the integrability of the general anharmonic
oscillator equation \cite{Eu},  three distinct classes of general
solutions of the highly nonlinear second order ordinary differential
equation
\be
\frac{d^{2}x}{dt^{2}}+f_{1}\left( t\right) \frac{dx}{dt}%
+f_{2}\left( t\right) x+f_{3}\left( t\right) x^{n}=0,
 \ee
 were presented in \cite{Ha3}. A class of exact solutions for the Li\'{e}nard type ordinary non-linear differential equation was obtained in \cite{Ha4}. In order to obtain the solutions, the second order Li\'{e}nard type differential equation is transformed into a second kind Abel type first order differential equation. With the use of an exact integrability condition for the Abel equation (the Chiellini lemma) \cite{Ha5}, the exact general solution of the Abel equation can be obtained, thus leading to a class of exact solutions of the Li\'{e}nard equation, expressed in a parametric form. The Chiellini integrability condition was also extended to the case of the generalized Abel equation. As an application of the integrability condition the exact solutions of some particular Li\'{e}nard type equations, including a generalized van der Pol type equation, were explicitly obtained.  The connection between dissipative nonlinear second order differential equations and the Abel equations, which in its first kind form have only cubic and quadratic terms, were considered in \cite{Rosu1}. By using the integrability criterion of Chiellini, the general solution of the corresponding integrable dissipative equations was obtained. Through a modified factorization method nonsingular parametric oscillators, which are Darboux related to the classical harmonic oscillator, and have periodic dissipative/gain features, have been identified \cite{Rosu2}. Note that the same method was also applied to the upside-down (hyperbolic) "oscillator". For these oscillators the obtained Darboux partners show transient underdamped features. Some basic results in the integrability of the Abel equations  were reformulated and expanded in \cite{Man4}.

 An interesting nonlinear oscillator, which is described by a Li\'{e}nard type equation, and which has a number of interesting characteristics, is
 \be\label{new11}
 \ddot{x}+kx\dot{x}+\frac{k^2}{9}x^3+\lambda _1x=0, \qquad k,\lambda _1={\rm constant}.
 \ee
 The mathematical properties of this equation have been studied extensively in \cite{new1}-\cite{new5}. Eq.~(\ref{new11}) can  also be considered as
a generalized Emden-type equation. It has the interesting property that for $\lambda _1>0$ nonisolated periodic orbits
can be explicitly obtained. These periodic orbits, which are of conservative Hamiltonian type,
have the interesting property that the frequency of the damped oscillations is completely independent of their amplitude. Moreover, they
continue to have the same value as the one corresponding to the linear harmonic oscillator \cite{new1}. In \cite{new2} it was shown that Eq.~(\ref{new11}) is integrable for $\lambda _1=0$. Eq.~(\ref{new11}) can be solved either explicitly, or by quadratures. It also turns out that it can be equivalently described in terms of some time-independent Hamiltonian functions. A particular case of Eq.~(\ref{new11}),
\be\label{new1}
\ddot{x}+3x\dot{x}+x^3+\omega ^2x=0,
\ee
was studied in \cite{new3}. Eq.~(\ref{new1}) has also the uncommon property that the frequency
of oscillation of the corresponding oscillator is independent of its amplitude. This property is specific for linear harmonic oscillators, and it is
not a general characteristic of nonlinear oscillator equations. An oscillator
 possessing the  property of independence of frequency from amplitude is also known as an isochronous
oscillator \cite{news1}.  The isochronicity property  of Eq.~(\ref{new1}) can be related to its main characteristic, namely
that it can be transformed through a nonlocal transformation  \cite{new3}
\be
U(t)=x(t)e^{\int_0^t{x\left(t'\right)dt'}},
\ee
to the linear harmonic oscillator equation,
\be
\ddot{U}(t)+\omega ^2U(t)=0.
\ee

A more general non-local transformation,
\be
U(t)=x(t)e^{\int_0^t{f\left(x\left(t'\right)\right)dt'}},
\ee
as applied to the linear harmonic oscillation equation,  generates a non-linear Li\'enard type differential equation of the form \cite{new3}
\be
\ddot{x}+\left(2f+xf'\right)\dot{x}+\left(f^2+\omega ^2\right)x=0,
\ee
where the prime denotes differentiation with respect to $x$. One can also consider  more general nonlocal transformations, defined as \cite{new3}
\be
U(t)=g\left(x(t)\right)e^{\int_0^t{f\left(x\left(t'\right)\right)dt'}},
\ee
leading to the equation
\be
\ddot{x}+\frac{g''}{g'}\dot{x}^2+\frac{gf'}{g'}\dot{x}+2f\dot{x}+\frac{g}{g'}\omega ^2+\frac{f^2g}{g'}=0,
\ee
where $g'=dg/dx$. A simple and efficient procedure to construct higher dimensional isochronous Hamiltonian
systems, called  the $\Omega $-modified procedure, was developed in \cite{news2}. By  using this procedure it was shown that a class of singular Hamiltonian systems is equivalent to constrained Newtonian systems.  The procedure was first proven for two dimensional dynamical systems, and then extended to the case of N-dimensional
systems. It was also shown that,  depending on the choice
of the system parameters, the considered systems admit two types of solutions: isochronous, and amplitude independent quasiperiodic solutions, respectively. The quantization of the nonlinear oscillators of type (\ref{new}) was considered in \cite{new4} and \cite{new5}, respectively.

A very powerful method of solving the differential equation,
\be\label{1.1}
 y''(x) +f_3(x, y)y’^2+f_2(x, y )y'+f_1(x, y) = 0,
\ee
where $f_i(x,y)$, $i=1,2,3$ are arbitrary functions, was proposed in \cite{Herb}. This method is as follows: assume that $f_3 =f_3(y)$ and  $f_2 =f_2(x)$. Then, provided that $f_1$ has the form
\be
f_1(x,y)=q(x)Z(y)-\exp\left[-2\int{f_2(x)dx}\right]C(y),
\ee
where $q(x)$ is arbitrary, and $Z'(y):=1-f_3Z$, $C'(y):=-\left(f_3+3/Z\right)C(y)$,  the general solution of Eq.~(\ref{1.1}) is $y=J(r,s)$, where $r$ and $s$ are variable independent solutions of the linear equation
\be
y''(x)+f_2(x)y'+q(x)y=0,
\ee
and the function $J$ is any solution of the system
\be
J_{rr} = s^2C(J) -f_3(J)J_r^2, \qquad J_{rs} = -rsC(J) -f_3(J)J_rJ_s,
 \ee
 \be
 J_{ss} = r^2C(J)-f_3(J)J_s^2, \qquad J_r=\left[Z(J)-sJ_s\right]/r.
 \ee
 In \cite{Das} it was shown that Eq.~{\ref{1.1}) can be transformed  into a first order equation, instead to a second order
linear equations. Using this transformation,  a first integral of Eq.~(\ref{1.1}) can be found. The results of \cite{Das} were generalized in \cite{Finch}, where a method for finding the complete first integral of the
Eq.(\ref{1.1}) was obtained, in the case when the coefficients $f_i$, $i=1,2,3$  fulfil a certain condition.

A method for obtaining a first integral of a Li\'{e}nard type equation of the form
\be
\ddot{x}+\frac{1}{x}\frac{d}{dx}\left[xf(x)\right]\dot{x}+\omega ^2x+\frac{f^2(x)}{x}=0,
 \ee
 where $f(x)$ is an arbitrary $C^1$ function, was introduced in \cite{Saw}.  The method is based on the introduction of a first order linear "generating equation" for a complex function $X=X\left(x,\dot{x}\right)$, which is equivalent to the equation of the linear harmonic oscillator, and by considering a "deformation" of this equation by means of the substitution $\dot{x}\rightarrow \dot{x}+f(x)$.

It is the purpose of the present paper to generalize the approach introduced in \cite{Saw}. First, by introducing, in analogy to the quantum mechanical treatment, two "creation" and "annihilation" functions $a=a\left(x,\dot{x},\omega\right)$ and $a^{+}=a^{+}\left(x,\dot{x},\omega\right)$, $\omega ={\rm constant}$, respectively, we define a generating function $X=X\left(a,a^{+}\right)$, satisfying a given particular differential equation. For a specific choice of $a$ and $a^{+}$ it follows that $x$ satisfies the simple linear harmonic oscillator equation, whose solution can be therefore obtained by using only the properties of the functions $a$ and $a^{+}$. By introducing a general "deformation" of the space-phase coordinates $\left(x,\dot{x}\right)$ so that
\be\label{1e}
x\rightarrow \tilde{x}= x+g\left(t,x,\dot{x}\right),
\ee
and
\be\label{2e}
\dot{x}\rightarrow \dot{\tilde{x}}= \dot{x}+f\left(t,x,\dot{x}\right),
\ee
with $f$ and $g$ being arbitrary $C^1$ functions of $t$, $x$, and $\dot{x}$, we construct the corresponding deformations of the functions $a\left(x,\dot{x},\omega\right)\rightarrow \tilde{a}\left(\tilde{x},\dot{\tilde{x}},\omega\right)$ and $ a^{+}\left(x,\dot{x},\omega\right)\rightarrow \tilde{a}^{+}\left(\tilde{x},\dot{\tilde{x}},\omega\right)$, respectively.

By imposing the condition that $\tilde{X}=\tilde{X}\left(\tilde{a},\tilde{a}^{+}\right)$ satisfies the same equation as in the case of the harmonic oscillator, a large class of non-linear Li\'{e}nard, or generalized Li\'{e}nard, type differential equations can be generated. These equations represent a non-linear extension of the harmonic oscillator equation, and their first integral can be easily obtained. Therefore the exact solution of several distinct classes of ordinary second order non-linear differential equations can be obtained in an exact analytical form.

The present paper is organized as follows. In Section~\ref{sect2} we review the problem of the linear harmonic oscillator, by using methods from quantum mechanics, and applying them to the classical case. The solution generating first order differential equation, and its solution are explicitly presented. In Section~\ref{sect3} we consider the phase space deformation of the coordinates $x$ and $\dot{x}$. The resulting differential equations, and their first integrals are explicitly presented. The exact solution of some non-linear differential equations are also obtained. We present a number of  possibilities of further generalizing our results in Section~\ref{sectn}. Some physical and engineering applications are briefly analyzed in Section~\ref{new}. We discuss and conclude our results in Section~\ref{sect4}.

\section{The linear harmonic oscillator}\label{sect2}

The equation of motion of the linear harmonic oscillator is given by
\be\label{1osc}
\ddot{x}(t)+\omega ^2 x(t)=0,\omega ={\rm constant}.
\ee
Based on the analogy with quantum mechanics \cite{Merz}  we introduce two complex "creation" and "annihilation" functions $a(t)$ and $a^{+}(t)$,  defined as
\be
a(t):=\dot{x}(t)-\mathrm{i}\omega x(t),
\ee
and
\be
a^{+}(t)\equiv a^{*}(t):= \dot{x}(t)+\mathrm{i}\omega x(t)
\ee
respectively, where $*$ denotes complex conjugation, and $x(t)$ is a solution of Eq.~(\ref{1osc}). Due to their definition, and the equation of motion of the harmonic oscillator, the functions $a$ and $a^{+}$ have the properties
\be
H(t)=a(t)a^{+}(t)=\dot{x}^2(t)+\omega ^2x(t)^2,
\ee
and
\be\label{cons}
\frac{dH(t)}{dt}=\frac{d}{dt}\left[a(t)a^{+}(t)\right]=\dot{a}(t)a^{+}(t)+a(t)\dot{a}^{+}(t)=0,
\ee
respectively. We introduce now the solution generating function $X$ defined as
\be
X(t)=\frac{a(t)}{a^{+}(t)}=\frac{\dot{x}(t)-\mathrm{i}\omega x(t)}{\dot{x}(t)+\mathrm{i}\omega x(t)}.
\ee

It follows that Eq.~(\ref{1osc}) is equivalent with the solution generating equation \cite{Saw},
\be\label{2}
\dot{X}(t)=-2\mathrm{i}\omega X(t).
\ee
Indeed, Eq.~(\ref{2}) gives
\be
\dot{a}a^{+}-a\dot{a}^{+}=-2\mathrm{i}\omega aa^{+},
\ee
or, with the use of Eq.~(\ref{cons}),
\be
\dot{a}=-\mathrm{i}\omega a,
\ee
from which Eq.~(\ref{1osc}) can be immediately obtained.

Eq.~(\ref{2}) has the general solution
\be\label{Xs}
X(t)=e^{-2\mathrm{i} \left(\omega t+\alpha\right)},
\ee
where $\alpha $ is an arbitrary constant of integration. Therefore from
\be
X(t)=\frac{a(t)}{a^{+}(t)}=\frac{\dot{x}(t)-\mathrm{i}\omega x(t)}{\dot{x}(t)+\mathrm{i}\omega x(t)}=e^{-2\mathrm{i} \left(\omega t+\alpha\right)},
\ee
 we obtain the first order linear differential equation
\be
\dot{x}(t)=\omega \cot \left(\omega t+\alpha \right)x(t),
\ee
with the general solution $x(t)=A\sin\left(\omega t+\alpha\right)$, where $A$ is an arbitrary constant of integration. Therefore we have obtained the general solution of the equation of motion of the linear harmonic oscillator by using the solution generating Eq.~(\ref{2}), and the definitions of the "creation" and "annihilation" functions $a(t)$ and $a^{+}(t)$, respectively.  In \cite{Rosu3} and \cite{Rosu4} an alternative factorization of the Hamiltonian of the quantum harmonic oscillator was considered. This factorization leads to the construction  of a specific two-parameter self-adjoint operator, from which the standard harmonic oscillator equation can be recovered in certain limits of the model parameters.

In the case $x=0$, $\dot{x}\neq 0$, we have $X(t)=1$, $a=a^{+}$. Then, from Eq.~(\ref{Xs}) it follows that the oscillating system reaches the position  $x=0$
at time moments given by
\be
t_n=\frac{n\pi-\alpha}{\omega}, \qquad n\in \mathbb{N}.
\ee

\section{Generalizing the linear harmonic oscillator equation by deforming the function $a(t)$}\label{sect3}

In the present Section we will generalize the form of the functions $a$ and $a^{+}$, and consequently of the solution generating function $X$, by "deforming" the phase - space coordinates $x$ and $\dot{x}$ according to Eqs.~(\ref{1e}) and (\ref{2e}). The deformation of the phase space coordinates will induce a deformation of  the "creation" and "annihilation" functions $a\left(x,\dot{x}\right)$ and $a^{+}\left(x,\dot{x}\right)$ of the linear harmonic oscillator. Therefore under the deformation of the coordinates the  "creation" function $a$ transforms as
\be
a\left(x,\dot{x},\omega\right)\rightarrow \tilde{a}\left(x+g\left(t,x,\dot{x}\right),\dot{x}+f\left(t,x,\dot{x}\right),\omega\right).
\ee

We will continue to  assume that the "deformed" operator $\tilde{X}$, given by
 \be
 \tilde{X}=\frac{\tilde{a}}{\tilde{a}^{+}}=\frac{\dot{x}(t)+f\left(t,x,\dot{x}\right)-\mathrm{i}\omega\left[x(t)+g\left(t,x,\dot{x}\right)\right]}{\dot{x}(t)+f\left(t,x,\dot{x}\right)-\mathrm{i}\omega\left[x(t)+g\left(t,x,\dot{x}\right)\right]},
 \ee
  satisfies the solution generating equation Eq.~(\ref{2}) of the simple harmonic oscillator,
  \be\label{n1}
  \dot{\tilde{X}}+2\mathrm{i}\omega \tilde{X}=0,
  \ee
  with the general solution given by
  \be
  \tilde{X}(t)=e^{-2\mathrm{i}\left(\omega t+\alpha\right)}.
  \ee
  The equation
  \be
  \tilde{a}=\tilde{a}^{+}e^{-2\mathrm{i}(\omega t+\alpha)},
  \ee
 gives the first integral of Eq.~(\ref{n1}). Let $F\left(t,x,\dot{x},\ddot{x}\right)=0$ a second order ordinary differential equation. A relation of the form $\Phi \left(t,x,\dot{x}\right)=C$, where $C$ is a constant, and $x$ is a solution of the equation $F\left(t,x,\dot{x},\ddot{x}\right)=0$, is called a first integral of $F\left(t,x,\dot{x},\ddot{x}\right)=0$ \cite{Ibr}.

 The procedure based on the "creation", "annihilation" and generating functions  allows us to obtain the first integrals, and therefore the exact solutions of a large class of ordinary non-linear differential equations, including   Li\'{e}nard and generalized Li\'{e}nard type equations. The most general deformation of the function $a(t)$ involves the presence of two arbitrary $C^1$ functions $f$ and $g$. The case $f\left(t,x,\dot{x}\right)\equiv 0$, $g\left(t,x,\dot{x}\right)\equiv 0$ corresponds to the case of the simple harmonic oscillator. For $g\left(t,x,\dot{x}\right)\equiv 0$, and $f\left(t,x,\dot{x}\right)\equiv x$ we reobtain the equation presented in \cite{Saw}.

 If the function $g\left(t,x,\dot{x}\right)$ satisfies the conditions
 \be
 x+g\left(t,x,\dot{x}\right)=0, \dot{x}+f\left(t,x,\dot{x}\right)\neq 0,
 \ee
 similarly to the case of the simple harmonic oscillator it follows that $\tilde{X}=1$, and the non-linear system passes through the point $x+g\left(t,x,\dot{x}\right)=0$ at the same time intervals as the simple harmonic oscillator passes through the point $x=0$.

 In the following we will assume that $f$ and $g$ are independent $C^1$ functions of the independent coordinate $t$, of the function $x$, and of $\dot{x}$, respectively.  By adopting specific forms for the functions $f$ and $g$ we will investigate the corresponding classes of equations, and their solutions.

\subsection{$a(t)=\dot{x}(t)+f(t)-\mathrm{i}\omega \left[x(t)+g(t)\right]$}

As a first case for the generalization  of the equation of the linear harmonic oscillations we consider the equation obtained by deforming the phase space coordinates $x(t)$ and $\dot{x}(t)$ so that $x(t)\rightarrow x(t)+g(t)$ and $\dot{x}(t)\rightarrow \dot{x}(t)+f(t)$, where $f(t)$ and $g(t)$ are  arbitrary $C^1$ functions  of time. Therefore the function $a(t)$ becomes
\be
a(t)=\dot{x}(t)+f(t)-\mathrm{i}\omega \left[x(t)+g(t)\right].
\ee
Then the solution generating equation Eq.~(\ref{2}) is equivalent with the second order non-linear differential equation
\bea
&&\left[ x(t)+g(t)\right] \ddot{x}(t)+\left[ f(t)-\dot{g}(t)\right] \dot{x}%
(t)+\omega ^{2}\left[ x(t)+g(t)\right] ^{2}+\nonumber\\
&&\dot{f}(t)\left[ x(t)+g(t)\right]
+\left[ f(t)-\dot{g}(t)\right] f(t)=0.  \label{4_1}
\eea
Eq.~(\ref{4_1}) has the first integral,
\be
\dot{x}=\omega \cot \left(\omega t+\alpha \right)\left[x(t)+g(t)\right]-f(t),
\ee
 which follows from $X(t)=a(t)/a^{+}(t)=e^{-2\mathrm{i} \left(\omega t+\alpha\right)}$.

 Therefore the general solution of Eq.~(\ref{4_1}) is given by
 \be
 x(t)=\sin\left(\omega t+\alpha \right)\left[A+\int{\frac{\omega \cot \left(\omega t+\alpha \right)g(t)-f(t)}{\sin\left(\omega t+\alpha \right)}dt}\right],
 \ee
where $A$ is an arbitrary constant of integration.

\subsubsection{ Particular case 1: $g(t)\equiv 0$}

If $g(t)\equiv 0$, Eq.~(\ref{4_1}) takes the form
\be\label{4}
x(t)\ddot{x}(t)+f(t)\dot{x}(t)+\omega ^{2}x^{2}(t)+\dot{f}(t)x(t)+f^{2}(t)=0,
\ee
with the first integral given by
\be
\dot{x}(t)=\omega \cot\left(\omega t+\alpha \right)-f(t).
\ee
Therefore the general solution  of Eq.~(\ref{4}) is given by
\be\label{5}
x(t)=\sin\left(\omega t+\alpha\right)\left[A-\int{\frac{f(t)}{\sin\left(\omega t+\alpha\right)}dt}\right].
\ee

As a simple application of the solution given by Eq.~(\ref{5}) we consider
the case $f(t)=f_{0}\sin \left( \omega t+\alpha \right) $, with  $f_{0}$
constant. Then Eq.~(\ref{4}) takes then the form
\begin{eqnarray}
&&x(t)\ddot{x}(t)+f_{0}\sin \left( \omega t+\alpha \right) \dot{x}(t)+\omega
^{2}x^{2}(t)+f_{0}\omega \cos \left( \omega t+\alpha \right) x(t)+  \nonumber
\\
&&f_{0}^{2}\sin ^{2}\left( \omega t+\alpha \right) =0,
\end{eqnarray}%
with the general solution given by
\begin{equation}
x(t)=\left( A-f_{0}t\right) \sin \left( \omega t+\alpha \right) .
\end{equation}

\subsubsection{Particular case 2: $f(t)\equiv 0$}

If $f(t)\equiv 0$, Eq.~(\ref{4_1}) reduces to
\begin{equation}
\left[ x(t)+g(t)\right] \ddot{x}-\dot{g}(t)\dot{x}(t)+\omega ^{2}\left[
x(t)+g(t)\right] ^{2}=0,  \label{g(t)}
\end{equation}%
with the general solution given by
\begin{equation}
x(t)=\sin (\omega t+\alpha )\left[ A+\int {\frac{\omega \cos \left( \omega
t+\alpha \right) }{\sin ^{2}\left( \omega t+\alpha \right) }g(t)dt}\right] .
\end{equation}%
If the function $g(t)$ has the simple functional form $g(t)=g_{0}\sin
^{n}\left( \omega t+\alpha \right) $, $g_{0}=\mathrm{constant}$, $n\in \mathbb{N}>1$, Eq.~(\ref{g(t)}) takes the form
\bea\label{gp}
&&\left[ x(t)+g_{0}\sin ^{n}\left( \omega t+\alpha \right) \right] \ddot{x}%
(t)-\omega ng_{0}\sin ^{n-1}\left( \omega t+\alpha \right) \cos \left(
\omega t+\alpha \right) \dot{x}(t)+\nonumber\\
&&\omega ^{2}\left[ x(t)+g_{0}\sin
^{n}\left( \omega t+\alpha \right) \right] ^{2}=0,\qquad n \in \mathbb{N}>1.
\eea
The general solution of Eq.~(\ref{gp}) is given by
\begin{equation}
x(t)=\sin (\omega t+\alpha )\left[ A+g_{0}\frac{\sin ^{n-1}\left( \omega
t+\alpha \right) }{n-1}\right] ,\qquad n \in \mathbb{N}>1.
\end{equation}

\subsection{$a(t)=\dot{x}+f(x)-\mathrm{i}\omega \left[x+g(x)\right]$}
As a second case of the deformation of the function $a$ we consider that $a(t)$ is of the form
\be
a(t)=\dot{x}+f(x)-\mathrm{i}\omega \left[x+g(x)\right],
\ee
where $f(x)$ and $g(x)$ are arbitrary $C^1$ functions of $x$. The differential equation satisfied by $x(t)$ can be obtained from Eq.~({\ref2}) as
\bea\label{jg}
&&\left[ x+g(x)\right] \ddot{x}-g^{\prime }(x)\dot{x}^{2}+\left\{ f^{\prime
}(x)\left[ x+g(x)\right] -f(x)\left[ g^{\prime }(x)-1\right] \right\} \dot{x}%
+\nonumber\\
&&\omega ^{2}\left[ x+g(x)\right] ^{2}+f^{2}(x)=0,
\eea
where a prime denotes the derivative with respect to the argument of the functions $f$ and $g$, respectively. Eq.~(\ref{jg}) has the first integral
\be\label{jpg1}
\dot{x}=\omega \cot \left(\omega t+\alpha \right)\left[x+g(x)\right]-f(x).
\ee

\subsubsection{Particular case 3: $g(x)=\left(\beta -1\right)x$, $f(x)=-\gamma x+\delta x^n$}

As an example of the application of the previous results we assume that the functions $g(x)$ and $f(x)$ have the functional forms
\be
g(x)=\left(\beta -1\right)x, \qquad f(x)=-\gamma x+\delta x^n,
\ee
where $n, \beta, \gamma, \delta ={\rm constants}$, and $n\in \mathbb{N}>1$.
With these choices Eq.~(\ref{jg}) takes the form
\bea\label{appl}
&&\beta  x \ddot{x}+(1-\beta ) \dot{x}^2+ \left[ (\beta  (n-1)+2)\delta  x^n-2 \gamma  x\right]\dot{x}-2 \gamma  \delta
   x^{n+1}+\delta ^2 x^{2 n}+\nonumber\\
   &&x^2
   \left(\beta ^2 \omega ^2+\gamma ^2\right)=0,\qquad  n\in \mathbb{N}>1.
\eea

The first integral Eq.~(\ref{jpg1}) becomes
\be
\dot{x}=\left[\beta \omega \cot(\omega t+\alpha )+\gamma \right]x-\delta x^n.
\ee
Therefore the general solution of Eq.~(\ref{appl}) is given by
\bea
x(t)&=&\sin ^{\beta \omega}\left(\omega t+\alpha\right)e^{\gamma t}\left[A-\delta \int{\sin^{(n-1)\beta \omega }(\omega t+\alpha )e^{(n-1)\gamma t}dt}\right]^{1/(1-n)}, \nonumber\\
&&n\in \mathbb{N}>1.
\eea

\subsubsection{Particular case 4: $g(x)\equiv 0$}

In the case $g(x)\equiv 0$, Eq.~(\ref{jg}) reduces to
\be\label{j}
\ddot{x}+\frac{1}{x}\left(\frac{d}{dx}xf(x)\right)\dot{x}+\omega ^2x+\frac{f^2(x)}{x}=0,
\ee

Eq.~(\ref{j}) is a standard Li\'{e}nard type equation. The first integral of this equation can be obtained from the condition $a(t)=a^{+}(t)e^{-2i\left(\omega t+\alpha \right)}$ as
\be\label{j1}
\dot{x}(t)=\omega \cot(\omega t+\alpha)x(t)-f(x).
\ee

Equations of the form (\ref{j}) and (\ref{j1}) were considered in \cite{Saw}, where it was shown that Eq.~(\ref{j}) can be exactly solved if the function $f(x)$ is of the form \cite{Saw}
\be
f(x)=-\beta x+\delta x^n,
 \ee
 with $n$, $\beta $ and $\delta $ constants. By assuming that
 \be
 f(x)=\mu x^2+\nu ,
 \ee
 where $\mu $ and $\nu $ are constants, Eq.~(\ref{j}) becomes
\be\label{nlin}
\ddot{x}+\frac{3\mu x^2+\nu}{x}\dot{x}+\left(\omega ^2+2\mu \nu\right)x+\mu ^2x^3+\frac{\nu ^2}{x}=0,
\ee
with the first integral given by
\be\label{ric}
\dot{x}=\omega \cot\left(\omega t+\alpha\right)-\mu x^2-\nu.
\ee
Eq.~(\ref{ric}) is a Riccati type differential equation \cite{Ha}. With the help of the transformation
\be
x=\frac{1}{\mu }\frac{\dot{v}}{v},
\ee
where $v$ is a new  $C^2$ function, Eq.~(\ref{ric}) can be transformed to a linear second order differential equation,
\be\label{lin1}
\ddot{v}=\omega \cot\left(\omega t+\alpha \right)\dot{v}-\mu \nu.
\ee
By introducing a new independent variable $\tau $ defined as
\be
\tau= -\omega \cos \left(\omega t+\alpha\right),
\ee
Eq.~(\ref{lin1}) takes the form
\be\label{ef}
\frac{d^2v}{d\tau ^2}+\frac{\mu \nu}{1-\tau ^2/\omega ^2}v=0.
\ee
Therefore the problem of solving the second order non-linear differential equation Eq.~(\ref{nlin}) has been reduced to the problem of solving the linear second order differential equation Eq.~(\ref{ef}). The general solution of Eq.~(\ref{ef}) is given by
\bea
v(\tau )&=& C_1 \, _2F_1\left(-\frac{1}{4} \sqrt{4 \mu  \nu  \omega
   ^2+1}-\frac{1}{4},\frac{1}{4} \sqrt{4 \mu  \nu  \omega
   ^2+1}-\frac{1}{4};\frac{1}{2};\frac{\tau ^2}{\omega ^2}\right)+\nonumber\\
 &&  \frac{C_2}{\omega}i  \tau
   \, _2F_1\left(\frac{1}{4}-\frac{1}{4} \sqrt{4 \mu  \nu  \omega ^2+1},\frac{1}{4}
   \sqrt{4 \mu  \nu  \omega ^2+1}+\frac{1}{4};\frac{3}{2};\frac{\tau ^2}{\omega
   ^2}\right), \nonumber\\
   \eea
   where $C_1$ and $C_2$ are arbitrary constants of integration.

   It is important to note that for $\nu =0$ Eq.~(\ref{nlin}) reduces to the Modified Emden Equation Eq.~(\ref{new}), whose properties have been studied in detail in \cite{new1}-\cite{new3} and \cite{new4,new5}.

   \subsubsection{Particular case 5: $f(x)\equiv 0$}

   In the case $f(x)\equiv 0$, Eq.~(\ref{jg}) becomes
\be\label{jpg2}
\ddot{x}-\frac{g'(x)}{x+g(x)}\dot{x}^2+\omega ^2\left[x+g(x)\right]=0,
\ee
with the first integral given by
\be
\dot{x}=\omega \cot \left(\omega t+\alpha \right)\left[x+g(x)\right].
\ee
Therefore the general solution of Eq.~(\ref{jpg2}) is given by
\be
\exp{\left[\int{\frac{dx}{x+g(x)}}\right]}=A\sin(\omega t+\alpha ).
\ee
In the case $g(x)=g_0x^n$, $g_0$, $n$ constants, Eq.~(\ref{jpg2}) takes the form
\be
\ddot{x}-\frac{g_0nx^{n-1}}{1+g_0x^{n-2}}\dot{x}^2+\omega ^2x+\omega ^2g_0x^n=0,
\ee
with the general solution given by
\be
\frac{x}{\left(1+g_0x^{n-1}\right)^{1/(n-1)}}=A\sin(\omega t+\alpha ).
\ee

\subsection{$a(t)=\dot{x}(t)+f\left(\dot{x}\right)-\mathrm{i}\omega \left[x(t)+g\left(\dot{x}\right)\right]$}

With the choice
\be
a(t)=\dot{x}(t)+f\left(\dot{x}\right)-\mathrm{i}\omega \left[x(t)+g\left(\dot{x}\right)\right],
\ee
where $f\left(\dot{x}\right)$ and $g\left(\dot{x}\right)$ are arbitrary functions of $\dot{x}$, the solution generating equation Eq.~(\ref{2}) is equivalent to the following non-linear generalized Li\'{e}nard type equation,
\begin{eqnarray}\label{jg4}
&&\left\{ \left[ f^{\prime }\left( \dot{x}\right) +1\right] g\left( \dot{x}%
\right) +x\left[ f^{\prime }\left( \dot{x}\right) +1\right] -\left[ f\left(
\dot{x}\right) +\dot{x}\right] g^{\prime }\left( \dot{x}\right) \right\}
\ddot{x}+
f\left( \dot{x}\right) ^{2}+\nonumber\\
&&\dot{x}f\left( \dot{x}\right) +\omega ^{2}%
\left[ g\left( \dot{x}\right) +x\right] ^{2}=0.
\end{eqnarray}
Eq.~(\ref{jg4}) has the first integral
\be
\dot{x}=\omega \cot\left(\omega t+\alpha\right)\left[x+g\left(\dot{x}\right)\right]-f\left(\dot{x}\right).
\ee

\subsubsection{Particular case 6:  $g(\dot{x})\equiv 0$}

In the particular case $g(\dot{x})\equiv 0$, we obtain the differential equation
\be\label{g3}
 \left[f'\left(\dot{x}\right)+1\right]\ddot{x}+\frac{f^2\left(\dot{x}\right)}{x}+\frac{\dot{x} f\left(\dot{x}\right)}{x}+\omega ^2 x=0.
\ee
 The general solution of Eq.~(\ref{g3}) can be obtained by using the first integral
\be
\dot{x}=\omega \cot (\omega t+\alpha )x-f\left(\dot{x}\right).
\ee

As an example we consider that the function $f\left(\dot{x}\right)$ is of the form
\be
f\left(\dot{x}\right)=-\frac{3}{4}\dot{x}+b,
\ee
with $b$ an arbitrary constant. For this choice of $f\left(\dot{x}\right)$, Eq.~(\ref{g3}) takes the form
\be
\frac{1}{4}  \ddot{x}-\frac{b}{2} \frac{\dot{x}}{x}-\frac{3}{16} \frac{\dot{x}^2}{x}+\frac{b^2}{x}+\omega ^2 x=0,
\ee
with the general solution given by
\bea
x(t)&=&\frac{1}{3 \omega }\Bigg\{2 b \sin \left[2 (\omega t +\alpha  )\right]+8 b \sin ^3(\omega t +\alpha ) \cos (\omega t +
\alpha )+\nonumber\\
&&3 c_1 \omega  \sin ^4(\omega t +\alpha  )\Bigg\},
\eea
where $c_1$ is an arbitrary constant of integration.

\subsubsection{Particular case 7: $f\left(\dot{x}\right)\equiv 0$}

In the case $f\left(\dot{x}\right)\equiv 0$, we obtain
\be\label{j4}
 \left[g\left(\dot{x}\right)+x-\dot{x} g'\left(\dot{x}\right)\right]\ddot{x}+\omega ^2 \left[g\left(\dot{x}\right)+x\right]^2=0,
   \ee
   and
   \be
\dot{x}=\omega \cot (\omega t+\alpha )\left[x+g\left(\dot{x}\right)\right],
\ee
respectively. With $g\left(\dot{x}\right)=c\dot{x}$, Eq.~(\ref{j4}) becomes
\be
 \ddot{x}+c^2 \omega ^2 \frac{\dot{x}^2}{x}+2 c \omega ^2  \dot{x}+\omega ^2 x=0,
\ee
with the general solution given by
\be
x(t)=A e^{-\frac{c \omega  (\alpha +t \omega )}{c^2 \omega ^2+1}} \left[|c \omega  \cos (\omega t+\alpha )-\sin (\omega t+\alpha )|\right]^{\frac{1}{c^2 \omega ^2+1}},
\ee
where $A$ is an arbitrary constant of integration.

 \subsection{$a(t)=\dot{x}+f\left(t,x,\dot{x}\right)-\mathrm{i}\omega \left[x+g\left(t,x,\dot{x}\right)\right]$}

 Finally, we consider the general deformation of the space phase coordinates $x$ and $\dot{x}$ by assuming a general dependence of the functions $f$ and $g$ on all coordinates $t$, $x$ and $\dot{x}$, respectively. The analysis of this case is straightforward, and represents a generalization of all the particular cases of deformations previously considered.

 Therefore we summarize the main results of our paper by formulating the following

 {\bf Theorem.} Let $f=f\left(t,x,\dot{x}\right)$ and $g=g\left(t,x,\dot{x}\right)$ be two $C^1$ arbitrary functions of the independent variable $t$, of the dependent variable $x$, and of its derivative $\dot{x}$, respectively. Then the  second order ordinary differential equation for $x$,
 \bea
 && \Bigg\{\left[\frac{\partial f\left(t,x,\dot{x}\right)}{\partial \dot{x}}+1\right]
   g\left(t,x,\dot{x}\right)+x \left[\frac{\partial f\left(t,x,\dot{x}\right)}{\partial \dot{x}}+1\right]-\left[f\left(t,x,\dot{x}\right) +\dot{x}\right]\times\nonumber\\
   &&\frac{\partial g\left(t,x,\dot{x}\right)}{\partial \dot{x}}
   \Bigg\}\ddot{x}+
 \Bigg\{\frac{\partial f\left(t,x,\dot{x}\right)}{\partial x} \left[g\left(t,x,\dot{x}\right)+x\right]-f\left(t,x,\dot{x}\right)
   \left[\frac{\partial g\left(t,x,\dot{x}\right)}{\partial x}-1\right]-\nonumber\\
   &&\frac{\partial g\left(t,x,\dot{x}\right)}{\partial t}\Bigg\}\dot{x}+
   \left[g\left(t,x,\dot{x}\right)+x\right] \left\{\frac{\partial f\left(t,x,\dot{x}\right)}{\partial t}+
   \omega ^2\left[
   g\left(t,x,\dot{x}\right)+ x\right]\right\}-\nonumber\\
   &&f\left(t,x,\dot{x}\right) \frac{\partial g\left(t,x,\dot{x}\right)}{\partial t}+
   f^2\left(t,x,\dot{x}\right)-
   \dot{x}^2
   \frac{\partial g\left(t,x,\dot{x}\right)}{\partial x}=0,
   \eea
 has the first integral
 \be
 \dot{x}(t)=\omega \cot\left(\omega t+\alpha\right)\left[x(t)+g\left(t,x,\dot{x}\right)\right]-f\left(t,x,\dot{x}\right).
 \ee

 \section{Further generalizations}\label{sectn}

 In the present Section we briefly consider the possibility of extending the method presented in this paper to the case of other types of differential equations. This can be done by generalizing the solution generating equation Eq.~(\ref{2})
to other integrable cases. For example, the solution generating exactly integrable equation
\be\label{61}
\dot{X}(t)=-2\mathrm{i}\omega X(t)-\mathrm{i}bX^2(t)-\mathrm{i}b,\qquad b={\rm constant}\neq \omega,
\ee
generates, for
\be
X(t)=\frac{\dot{x}(t)-\mathrm{i}\omega x(t)}{\dot{x}(t)+\mathrm{i}\omega x(t)},
\ee
the non-linear differential equation
\be\label{fin0}
 \ddot{x}(t)+\frac{b}{\omega } \frac{\dot{x}(t)^2}{x(t)}-\omega  (b-\omega ) x(t)=0.
\ee

The general solution of Eq.~(\ref{61}) is given by
\be
X(t)=-\frac{\omega +\mathrm{i} \sqrt{b^2-\omega ^2  } \tanh \left[\sqrt{b^2-\omega ^2 } (\alpha +t)\right]}{b},
\ee
and therefore Eq.~(\ref{fin0}) has the first integral
\be
\dot{x}(t)=\omega\frac{  \sqrt{b^2-\omega ^2 } \tanh \left[\sqrt{b^2-\omega ^2} (t+\alpha )\right]+\mathrm{i} (b-\omega )}{\mathrm{i} \sqrt{b^2-\omega ^2} \tanh
   \left(\sqrt{b^2-\omega ^2 } (t+\alpha )\right)+b+\omega }x(t).
\ee

By deforming the phase space coordinates by introducing the functions $f$ and $g$, from Eq.~(\ref{61}) several classes of non-linear differential equations, as well as their first integrals, can be generated.

Another possibility of generalizing the previous formalism is to consider that the oscillation frequency of the oscillator is a function of time, $\omega =\omega (t)$, that is, to start with the generating function
\be\label{df}
X(t)=\frac{\dot{x}(t)-\mathrm{i}\omega (t)x(t)}{\dot{x}(t)+\mathrm{i}\omega (t) x(t)}.
\ee

Then the generating equation $\dot{X}(t)+2i\omega (t)X(t)=0$ is equivalent with the differential equation
\be
\ddot{x}(t)-\frac{\dot{\omega (t)}}{\omega (t)}\dot{x}(t)+\omega ^2(t)x(t)=0,
\ee
which has the first integral
\be
\dot{x}(t)=\omega (t)\cot \left[\int{\omega \left(t'\right)dt'}+\alpha\right]x(t).
\ee

The time deformation of the phase space coordinates in Eq.~(\ref{df}), $x(t)\rightarrow x(t)+g(t)$, $\dot{x}(t)\rightarrow \dot{x}(t)+f(t)$  generates the differential equation
\bea
&&\left[g(t)+x(t)\right] \ddot{x}(t)+ \left[f(t)-\dot{g}(t)-\frac{g(t) \dot{\omega }(t)}{\omega (t)}-\frac{ \dot{\omega }
   (t)}{\omega (t)}x(t)\right]\dot{x}(t)+\nonumber\\
&& \left[\dot{f}(t)-\frac{f(t) \dot{\omega }(t)}{\omega (t)}+2 g(t) \omega ^2(t)\right]x(t)+ \omega ^2(t)x^2(t)-f(t) \dot{g}(t)-\nonumber\\
&&   \frac{f(t) g(t) \dot{\omega }(t)}{\omega (t)}+f^2(t)+g^2(t) \omega ^2(t)+g(t) \dot{f}(t)=0,
\eea
which has the first integral
\be
\dot{x}(t)=\omega (t)\cot \left[\int{\omega \left(t'\right)dt'}+\alpha \right]\left[x(t)+g(t)\right]-f(t).
\ee

The systematic study of these types of differential equations will be performed in a future study.

In the case of the linear harmonic oscillator the quantity $H=aa^{+}$ represents the conserved energy of the system. For the non-linear generalization of the harmonic oscillator considered in the present paper one can define the energy of the oscillator as
\be
H=aa^{+}=\l(\dot{x}+f\r)^2+\omega ^2\l(x+g\r)^2=\omega ^2\frac{(x+g)^2}{\sin^2\l(\omega t+\alpha\r)}.
\ee

The time variation of the energy is given by
\be
\frac{dH}{dt}=2\omega ^2\frac{\l(x+g\r)\l(\dot{g}-f\r)}{\sin^2\l(\omega t+\alpha\r)}.
\ee
It is interesting to note that the energy is conserved not only for $f=g\equiv0$, but also for deformations of the  phase space coordinates that satisfy the condition $\dot{g}=f$.

 \section{Applications in Physics and Engineering}\label{new}

 As we have already mentioned, Eq.~(\ref{nlin}) represents an integrable generalization of the equation of anharmonic oscillations Eq.~(\ref{new}), which arises in a large number of physical problems (see \cite{new3} and references therein). From a physical point of view Eq.~(\ref{nlin}) describes the motion of a point particles under the influence of dissipation, described by a term of the form $\left(3\mu x+\nu /x\right)\dot{x}$, and an exterior anharmonic force $\omega ^2x+\mu ^2x^3+\nu ^2/x$. By using the obtained results several generalizations of the Rayleigh equation \cite{Nay},
 \be
 \ddot{x}+F(\dot{x})+x=0,
 \ee
 describing non-linear systems with one degree of freedom,  admitting auto-oscillations, can also be solved exactly.

 In the following we will briefly present some physical and engineering applications of the methods introduced in the present paper in two areas of physics and engineering: travelling waves in reaction-convection-diffusion systems,  and the approximate solution of the equation describing large-amplitude
non-linear oscillations of a uniform cantilever beam carrying an intermediate lumped mass and rotary inertia, respectively.

 \subsection{Travelling wave solutions in reaction-convection-diffusion systems}

 From a mathematical point of
view reaction--convection--diffusion equations represents a particular class of second order non-linear partial differential equations, which play a very
important role in many areas of physics, chemistry, engineering and biology. A large number of distinct phenomena,
including heat transfer, combustion, reaction chemistry, fluid dynamics,
plasma physics, soil-moisture, foam drainage, crystal growth, biological
population genetics, cellular ecology, tumor growth, neurology, synergy etc. can be described mathematically by using reaction - convection - diffusion equations \cite{9cm} - \cite{Har}. The general class of reaction - convection - diffusion equations is given by \cite{9cm,b3}
\begin{equation}  \label{1}
\frac{\partial u}{\partial t}=\frac{\partial }{\partial x}\left[D(u)%
\frac{\partial u}{\partial x}\right]+B(u)\frac{\partial u}{%
\partial x}+Q(u),
\end{equation}
where $u=u(t,x)$, and the function $D(u)$ plays the role of the diffusion coefficient, $B(u)$
may be viewed as a nonlinear convective flux function, and $Q(u)$ is the
reaction function, respectively. Particular classes of reaction-convection-diffusion equations are the Fisher, or the logistic equation,  the Newell-Whitehead, or amplitude equation, the Zeldovich equation, and the Nagumo or the bistable
equation \cite{b3}.

Reaction-diffusion equations have usually the important property of admitting travelling wave solutions \cite{9cm}-\cite{Har}. Travelling wave phenomena have been observed, or produced experimentally, in a large number of natural reaction, convection and
diffusion processes, and they play an important role, from both theoretical and experimental point of view, in many fields of physics, chemistry, biology, and engineering. By introducing a phase variable $\xi =x-V_{f}t$, where $V_{f}={\rm constant}\geq
0$ is a constant wave velocity, and by assuming that $u=u\left(x-V_ft\right)$, Eq.~(\ref{1}) takes the form of a second
order non-linear differential equation of the form
\begin{equation}\label{101}
\frac{d^{2}u}{d\xi ^{2}}+\alpha (u)\left( \frac{du}{d\xi }\right) ^{2}+\beta
(u)\frac{du}{d\xi }+\gamma (u)=0,
\end{equation}%
where
\begin{equation}
\alpha (u)=\frac{d}{du}\ln D(u),\beta (u)=\frac{V_{f}+B(u)}{D(u)},\gamma (u)=%
\frac{Q(u)}{D(u)}.
\end{equation}%

Eq.~(\ref{101}) has the same mathematical form as Eq.~(\ref{jg}). Therefore it follows that if the reaction function $D(u)$, the convection function $B(u)$ and the reaction function $Q(u)$ can be represented in terms of two arbitrary function $f(u)$ and $g(u)$ as
\be
\alpha (u)=-\frac{g'(u)}{u+g(u)},
\ee
\be
\beta (u)=f'(u)-\frac{f(u)\left(g'(u)-1\right)}{u+g(u)},
\ee
\be
\gamma (u)=\omega ^2\left[u+g(u)\right]+\frac{f^2(u)}{u+g(u)},
\ee
then the general reaction-convection-diffusion equation Eq.~(\ref{101}) has the first integral
\be
\frac{du}{d\xi }=\omega \cot (\omega \xi+\alpha)\left[u+g(u)\right]-f(u).
\ee

Concentration-dependent diffusion, convection,  and  reaction coefficients are  used to model mass transfer
phenomena in polymer membranes \cite{57}, as well as water sorption in a polymer
matrix composite \cite{58,59}. Several functional forms for the concentration dependence
of the diffusion, convection and reaction functions have been proposed, and used to model various phenomena,  including the very simple form of a linear dependence of the diffusion function $D(u)=D_0\left(1+k_Du\right)$, where $D_0$ and $k_D$ are constants \cite{57}.

\subsubsection{The case $g(u)=(\beta -1)u$ and $f(u)=-\gamma u+\delta u^2$}

As an application of the above results on the exact integrability of the reaction-convection-diffusion equations, we consider the simple case
\be
g(u)=(\beta -1)u,
\ee
and
\be
f(u)=-\gamma u+\delta u^2,
\ee
respectively, where $\beta $, $\gamma $ and $\delta $ are constants. Thus we obtain
\be
\alpha (u)=\frac{1-\beta}{\beta u},
\ee
\be
\beta (u)=-\frac{2\gamma }{\beta }+\delta \left(1+\frac{2}{\beta }\right)u,
\ee
\be
\gamma (u)=\omega ^2\beta u+\frac{(-\gamma +\delta u)^2}{\beta },
\ee
giving
\be\label{D1}
D(u)=D_0u^{(1-\beta )/\beta },
\ee
\be\label{B1}
B(u)=D_0u^{(1-\beta )/\beta}\left[-\frac{2\gamma }{\beta }+\delta \left(1+\frac{2}{\beta}\right)u\right]-V_f,
\ee
\be\label{Q1}
Q(u)=D_0u^{(1-\beta )/\beta}\left[\omega ^2\beta u+\frac{\left(-\gamma +\delta u\right)^2}{\beta }\right],
\ee
where $D_0$ is an arbitrary constant of integration.

With this choice of the reaction, convection and diffusion functions, Eq.~(\ref{101}) has the first integral
\be
\frac{du}{d\xi}=\left[\beta \omega \cot (\omega \xi+\alpha)+\gamma \right]u-\delta u^2,
\ee
and has the exact travelling wave solution
\be
u(\xi)=\frac{\sin ^{\beta \omega }(\omega \xi +\alpha)e^{\gamma \xi }}{A-\delta \int{\sin^{\beta \omega }(\omega \xi +\alpha)e^{\gamma \xi}d\xi}}.
\ee
Hence we have obtained an exact travelling wave solution of the general reaction-convection-diffusion Eq.~(\ref{101}), under the assumptions of a specified form of $D(u)$, $B(u)$ and $Q(u)$. If the coefficients $\beta $ and $\omega $ satisfy the condition  $\beta \omega =1$, then the general solution of the reaction-convection-diffusion equation (\ref{101}) with diffusion, convection and reaction functions (\ref{D1})-(\ref{Q1}) is given by
\be
u(\xi)=\frac{\sin (\omega \xi +\alpha)e^{\gamma \xi }}{A-\delta e^{\gamma  \xi } \left[\gamma  \sin (\omega \xi   +\alpha  )-\omega  \cos (\omega \xi
     +\alpha )\right]/\left(\gamma ^2+\omega ^2\right)}.
\ee

\subsection{Large amplitude free vibrations of a uniform cantilever beam}

Many engineering structures can be modelled as a slender, flexible cantilever beam, which at an intermediate point along its span carries
a lumped mass with rotary inertia  \cite{Ham}.  In the linear theory
the vibrations of a cantilever beam are usually analyzed by introducing the approximation that assumes that the frequency of the free vibrations
is independent of the  amplitude of the motion, that is, that the system is isochronous. This approximation is valid only when the amplitude of the oscillations is
relatively small. However, since beam systems are slender and flexible, their flexural vibrations often have
relatively large amplitudes. Due to  their mathematical complexity, such nonlinear
problems do not allow an exact solution, and therefore approximate techniques must
be used for their study. In \cite{Ham}, to obtain the approximate solution to the period of oscillation of the cantilever beam the single-term harmonic balance method, and the two terms harmonic balance method was used. Several approximate or semi-analytical mathematical methods have been developed for the study of the free large amplitude nonlinear oscillations of a
slender, inextensible cantilever beam, like the homotopy
perturbation method \cite{p1},  modified
Lindstedt-Poincare methods \cite{p2,p3}, and the Optimal
Homotopy Asymptotic Method \cite{p4}. In \cite{p5} six different analytical
methods are applied to solve the dynamic model of the large amplitude non-linear
oscillation equation.

In the following we adopt the non-linear oscillation model introduced in \cite{Ham}. We consider a beam clamped at the base, and free at the tip.
The beam caries at an arbitrary
intermediate point $s=d$ along its span a lumped mass $M$ having a rotary inertia $J$. To simplify or mathematical model we introduce the conservative assumption that the beam is uniform, having constant length $l$, and mass $m$
per unit length, respectively. As compared to its length, the thickness of the
beam is assumed to be small. Moreover, in the following we ignore the effects of the rotary inertia and of the shearing deformations.
Also the beam is assumed to have the important property of being inextensible. Note that this implies that the length of the beam
neutral axis remains constant during the entire oscillatory motion. Then, with the help of the above assumptions,  the potential energy $V$, due to the bending of the beam, can be written as
\be
V=\frac{EIl}{2}\int_0^1{R^2\left(\xi,t\right)d\xi},
\ee
where $EI$ is  the modulus of flexural rigidity, $\xi =s/l$ is the dimensionless arc length, and $R(\xi, t)$ is the radius of curvature of the beam neutral axis. The kinetic energy $T$ of the beam is given by
\be
T=\frac{ml}{2}\int_0^1{\left(\dot{x}^2+\dot{y}^2\right)d\xi}+\frac{1}{2}M\left.\left(\dot{x}^2+\dot{y}^2\right)\right|_{\eta l}+\frac{1}{2}J\left.\dot{\theta}^2\right|_{\eta l},
\ee
where $\eta =d/l$, a dimensionless quantity, is the relative position parameter of the attached inertia element, and $\theta $ is the slope of the elastica, given by $sin\theta =dy/ds$. In order to simplify the mathematical analysis one expands first the kinetic and potential energies into power series, with the
nonlinear terms retained up to fourth order. Then, by looking for an approximate single mode solution  of the form $y(\xi,t)=\Phi (\xi )u(t)$, one obtains by simple calculations the discrete single mode, single coordinate cantilever beam Lagrangian as
\be\label{L}
L=\frac{ml}{2}\left(\alpha _1\dot{u}^2+\alpha _3\lambda ^2u^2\dot{u}^2-\beta _1^2\alpha _2u^2-\beta ^2\alpha _4\lambda ^2u^4\right),
\ee
where $\alpha _1$, $\alpha _2$, $\alpha _3$, $\alpha _4$, $\lambda $, and $\beta _1$ are constants. After a rescaling of the time coordinate and of the constants $\alpha _i$, $i=1,2,3,4$, $\lambda $ and $\beta $, we obtain the equation of motion corresponding to the Lagrangian (\ref{L}) as \cite{Ham}, \cite{p4}
\be
\ddot{u}+\alpha \frac{u}{1+\alpha u^2}\dot{u}^2+\omega ^2\left(\frac{u}{1+\alpha u^2}+\beta \frac{u^3}{1+\alpha u^2}\right)=0.
\ee

We write the above equation in the form
\be\label{119}
\ddot{u}+\alpha \frac{u}{1+\alpha u^2}\dot{u}^2+\omega ^2\left[u+\frac{u^3\left(\beta -\alpha \right)}{1+\alpha u^2}\right]=0,
\ee
where $\alpha =\alpha _3/\left(p^2\alpha _1\right)$ and $\beta =2\alpha _4/\left(p^2\alpha _2\right)$, where $p^2=\Omega /\beta $ is the dimensionless frequency, and $\Omega $ is the frequency of the assumed mode of the associated linear beam \cite{Ham}.
We try to obtain some approximate solutions of Eq.~(\ref{119}) by using the analogy between Eq.~(\ref{119}) and Eq.~(\ref{jpg2}). In order to exploit this analogy we first introduce the function $g(u)$, satisfying the differential equation
\be
  \frac{\alpha u}{1 + αu^2}= - \frac{g'(u)}{u+g},
\ee
with the general solution given by
\be
g(u)=\frac{1}{2} \left[\frac{2 \sqrt{\alpha } c_1+\sinh ^{-1}\left(\sqrt{\alpha }
   u\right)}{\sqrt{\alpha  \left(1+\alpha  u^2\right)}}-u\right],
\ee
where $c_1$ is an arbitrary constant of integration. In order to obtain an approximate solution of Eq.~(\ref{119}) we tentatively try to approximate the function
\be
h(u)=\frac{(\beta-\alpha)u^3}{\left(1+\alpha u^2\right)},
\ee
by $g(u)$. To see if such an approximation is valid, we expand both $g(u)$ and $h(u)$ in power series near $u=0$. We thus obtain
\be\label{122}
g(u)\approx c_1-\frac{1}{2}\alpha  c_1 u^2 -\frac{\alpha  u^3}{3}+\frac{3}{8} \alpha
   ^2 c_1 u^4+\frac{4 \alpha ^2 u^5}{15}-\frac{5}{16}\alpha ^3
   c_1 u^6 +O\left(u^7\right),
\ee
\be
h(u)\approx (\beta -\alpha )u^3 +\alpha \left(\alpha -  \beta \right)u^5 +O\left(u^7\right).
\ee
For $c_1=0$ Eq.~(\ref{122}) becomes
\be
g(u)\approx -\frac{\alpha  u^3}{3}+\frac{4 \alpha ^2 u^5}{15}+O\left(u^7\right).
\ee

The function $-g(u)$, as given by Eq.~(\ref{122}), is not a good approximation of $h(u)$  for arbitrary $\beta $. However, for the particular case $\beta=2/3 \alpha $ we obtain
\be
h(u) \approx −\alpha \frac{u^3}{3} + O(u^5),
\ee
\be\label{132n}
g(u) \approx −\alpha \frac{u^3}{3} + O(u^5).
\ee
Therefore for this particular case, and in this order of approximation,  Eqs.~(\ref{119}) and (\ref{jpg2}) are the same. Hence in Eq.~(\ref{132n}) the $\approx $ sign could be substituted by the $=$ sign.

Thus, by approximating $h(u)$ by $g(u)$, it follows that the solution of Eq.~(\ref{119}) can be represented in an integral form as
\be\label{125}
\exp\left[2 \int \frac{1}{\frac{\sinh ^{-1}\left(\sqrt{\alpha }
   u\right)}{\sqrt{\alpha  \left(\alpha  u^2+1\right)}}+u} \, du \right]\approx A\sin (\omega t+\alpha ).
\ee

Eq.~(\ref{125}) can be used as a starting point for obtaining some power series representations of the solutions of
the large-amplitude free vibrations of the considered
slender inextensible cantilever beam, which is
assumed to be undergoing planar flexural vibrations. On the other hand it is important to mention that Eq.~(\ref{125}) may not hold for arbitrary $\beta $.

From a physical point of view the terms proportional to $\alpha $ in Eq.~(\ref{119}) are of
inertia type. They are due to the kinetic energy of the axial motion, and they arise as a result of using the
approximation of the inextensibility condition. The first of these two non-linear terms
has a softening effect, while the second has a hardening effect \cite{Ham}. On the other hand the term proportional to $\beta $
in Eq.~(\ref{119}) is of a hardening static type, due to the potential energy stored
in bending \cite{Ham}. Therefore the condition $\beta =(2/3) \alpha $ can be interpreted as giving a specific condition relating the potential energy stored in the bending with the kinetic energy of the axial motion.

\section{Discussions and final remarks}\label{sect4}

In the present paper we have considered a class of differential equations that can be obtained by considering a deformation of the phase space coordinates of the linear harmonic oscillator, while keeping unchanged the solution generating equation of the harmonic oscillations. The devised method is an application of the quantum mechanical approach to the harmonic oscillations to the classical regime. The basic quantities used in the present approach are the "creation" and "annihilation" functions $a$ and $a^{+}$, which in quantum mechanics are
represented by Hermitic operators \cite{Merz}, and the solution generating equation, constructed with the help of these functions. By deforming the phase space of the harmonic oscillator by adding two arbitrary functions $f$ and $g$ to the position and velocity coordinates, respectively, the solution generating equation leads to several classes of dissipative, non-linear ordinary second order differential equations, representing a non-linear generalization of the harmonic oscillator. The non-linear generalizations of the linear oscillations  can be constructed by using a systematic approach based on the adopted form of the "deformation" and of the "annihilation" and "creation" functions. All the resulting equations have a first integral given by the first order differential equation
\begin{equation}
\frac{\dot{x}+f}{x+g}=\omega \cot\l(\omega t+\alpha\r).
\end{equation}

An interesting question, intrinsically related to the methods developed in the present paper, is the important question of the isochronous nature of the solutions, that is, the independence of the amplitude of the oscillation on its frequency. As was shown in \cite{new3}, all equations that can be transformed to the linear harmonic oscillation equation do have this property. On the other hand, for all the solutions obtained in the present paper the angular frequency $\omega $ of the oscillations of the "deformed" system continues to remain the same as that of the linear oscillation.
Therefore all the obtained solutions exhibits the property of amplitude independence of the frequency of oscillation, and the deformation of the phase space of the simple linear harmonic oscillator generates isochronous oscillating systems.

As some simple examples of applications of the obtained results in physics and engineering we have briefly discussed the possibility of obtaining exact travelling wave solutions of the general reaction-convection-diffusion systems, and approximate solutions for the nonlinear equation describing the large amplitude free vibrations of a uniform cantilever beam. The travelling wave equation for the general reaction-convection-diffusion systems belongs to the type of equations studied in the present paper, and as such it can be interpreted mathematically as being related to a phase space deformation of the harmonic oscillator equation. For variable diffusion, convection and reaction functions, satisfying some given relations, a large number of travelling wave solutions can be obtained. We have also presented a particular exact travelling wave solution, obtain by prescribing some specific functional forms for the functions $D$, $B$ and $Q$.

Most of the mathematical studies of the reaction--convection--diffusion equations have been done by using either asymptotic evaluations, or qualitative methods. The usual way to study nonlinear
reaction--diffusion equations similar to Eq.~(\ref{101}) consists of the analysis of the behavior of the system on a phase plane
$(du/d\xi, u)$. This approach,  useful for qualitative analysis, is insufficient for finding any exact solution, or for testing a numerical
one. Finding exact analytical solutions of the complicated reaction--convection--diffusion models can considerably simplify the process of comparison of the theoretical predictions with the experimental data, without the need of using numerical procedures.

Even it presents strong similarities with the differential equations studied in the present paper, the equation (\ref{119}) describing the nonlinear oscillations of uniform cantilever beams does not belong to the class of equations that can be obtained by deforming the linear harmonic oscillator equation. However, approximate semi-analytical solutions of the equation can be obtained under the assumption that the generalized anharmonic term, proportional to $u^3$, can be approximated by the same function $g$ that appears as the coefficient of the term $\dot{u}^2$ in the exact solution. Of course the validity of such an approach must be checked by performing a full comparison between the exact numerical solution, and the different order approximations obtained by using the procedure introduced in the present paper.

\section*{Acknowledgments}

We would like to thank to the three anonymous referees for comments and suggestions that helped us to significantly improve our manuscript. The authors are very grateful to Dr. Christian B\"ohmer for his careful reading of the manuscript.

\end{document}